\documentclass[aps,prx,twocolumn,showpacs,superscriptaddress,floatfix,amsmath,amssymb,longbibliography]{revtex4-2}

\usepackage[titletoc,toc,title]{appendix}

\usepackage{graphicx}
\usepackage{bm}
\usepackage{xcolor}
\usepackage{booktabs}
\usepackage{hyperref}
\usepackage{diagbox}
\usepackage{braket}
\usepackage{soul}
\usepackage[color=brown!60,textsize=scriptsize]{todonotes}
\renewcommand{\vec}[1]{\bm{#1}}
 \definecolor{dblue}{HTML}{517C99}
\definecolor{fountain}{HTML}{2197A9}
\definecolor{river}{HTML}{517C96}
\definecolor{lriver}{HTML}{dbe2ea}

\def\ie{\emph{i.e.,} }

\newcommand{\wqo}{$\overline{\omega}_{Q}$}
\begin{document}
	 \title{Orbital glass conceals missing  magnetic entropy in a  relativistic Mott insulator
  }

		\author{Ilija K. Nikolov}
	\affiliation{Department of Physics, Brown University, Providence, Rhode Island 02912-1843, USA}
	\author{Rong Cong}
	 \email[]{current affiliation National High Magnetic Field Laboratory,  Tallahassee, Florida \& equal contribution with the first author}
	\affiliation{Department of Physics, Brown University, Providence, Rhode Island 02912-1843, USA}
		\author{Adrien Rosuel}
	\affiliation{Department of Physics, Brown University, Providence, Rhode Island 02912-1843, USA}
	\author{Stephen Carr}
	\affiliation{Department of Physics, Brown University, Providence, Rhode Island 02912-1843, USA}
	\affiliation{Brown Theoretical Physics Center, Brown University, Providence, Rhode Island 02912-1843, USA.}
		\author{Ian R. Fisher}
	\affiliation{Department of Applied Physics and Geballe Laboratory for Advanced Materials, Stanford University, Stanford, California 94305, USA.} 
	\affiliation{Stanford Institute for Materials and Energy Sciences, SLAC National Accelerator Laboratory, 2575 Sand Hill Road, Menlo Park, California 94025, USA}
	\author{Dmitri E. Feldman}
	\affiliation{Department of Physics, Brown University, Providence, Rhode Island 02912-1843, USA}
	\affiliation{Brown Theoretical Physics Center, Brown University, Providence, Rhode Island 02912-1843, USA.}
	\author{Adrian Del Maestro}
	\affiliation{Department of Physics and Astronomy, University of Tennessee, Knoxville, TN 37996, USA}
	\affiliation{Min H. Kao Department of Electrical Engineering and Computer Science,
		University of Tennessee, Knoxville, TN 37996, USA}
	\author{Chandrasekhar Ramanathan}
	\affiliation{Department of Physics and Astronomy, Dartmouth College, Hanover, NH 03755, USA}
	\author{Vesna F. Mitrovi\'c}
    \email[]{corresponding author: vemi@brown.edu}
	\affiliation{Department of Physics, Brown University, Providence, Rhode Island 02912-1843, USA}
	

	\begin{abstract}
 \textbf{
     Coupling between different degrees of freedom (DOF) in an electronic material  leads to exotic phases of matter characterized by  complex  and competing order parameters  as well as emergent excitations.
Building  a microscopic understanding of these  order parameters  and their mutual relationship   is hindered by the fact that different orders often mask  each others'   response to conventional  
 experimental probes. 
		Here, we reveal how to disentangle responses from distinct orders 
		 that arise from the coupling between the spin and orbital DOF.  Our  method uses a phase sensitive technique that measures  ground state properties by independently resolving interactions of different symmetries.  
 This allows us to  directly detect an orbital glass state 
		 caused by  competing interactions  
                 in the $5d^1$ relativistic Mott insulator Ba$_2$NaOsO$_6$. 		We observe short-range orbital order up to~380~K and a dramatic increase of orbital dispersion near the magnetic phase transition. 
                This orbital dispersion generates  a  directional   ordering, \ie it forms an orbital nematic state which breaks the rotational symmetry of the crystal. We establish that the orbital nematic state  induces the magnetic ordering. 		The presence of this short-range orbital order 
                 well above the magnetic phase transition solves the long-standing puzzle of missing entropy in this material.  
  }
	\end{abstract}

	\maketitle
	
        Quantum materials can have complex electronic Hamiltonians with multiple unknown terms leading to different emergent phases of matter, each potentially characterized by an order parameter and its topology~\cite{Ramirez1994, OrbitPhysTMO, Pesin:2010aa, Santini2009, Hasan2010, Onoda2010, Lee2012, Chandra2013, Tsujimoto2014, WitczakKrempa2014,   2016AA, Wen2017, khomskii_orbital_2021,Khomskii_2022,Yin:2022aa,Chen:2024aa}. A prime example of such complexity arises in strongly-correlated electron systems with strong spin-orbit coupling (SOC)~\cite{Elliott1954, Pesin:2010aa, WitczakKrempa2014, Manchon2015, Soumyanarayanan2016,PhysRevResearch.2.013353}. While the single-particle picture along with perturbative corrections provides an unexpectedly useful  
      classification of  correlation physics  of many phenomena, \textit{e.g.,} nuclear structure~\cite{Bender2003,   Barrett2013, Hagen2014}, it fails in the presence of strongly interacting electrons and SOC whose complex interplay induces non-trivial states of matter. Their description remains  a grand challenge of condensed matter physics~\cite{Tsymbal2013,Rau2016,Pesin:2010aa,WitczakKrempa2014}. The emergent phases include quantum spin liquids~\cite{WitczakKrempa2014, Savary2016,Zhou2017}, cooperative Jahn-Teller effects~\cite{WitczakKrempa2014}, multipolar orderings~\cite{Harter2017, Takayama2021}, correlated topological semi-metals~\cite{Takayama2021}, and topological superconductivity~\cite{Rau2016, Wen2017}.
        Most experimental probes are sensitive to only a single aspect of an emergent phase, \ie a local order parameter~\cite{Gingras01,comin_broken_2015,Kung2015,Multipolarhidden19,Zhou:2020aa} that couples to the applied field or quasi-particles used for the measurement. 
        As such, it can be difficult to discern which  of the potentially competing  interactions drives an observed  phase transition, as well as their mutual relationships~\cite{Tonegawa:2014aa, Mitsumoto2020, PhysRevLett.78.2799,PhysRevB.75.195113,PhysRevResearch.2.013353,Gedik23}. Consequently, important  (hidden) order parameters and/or interactions  can  remain elusive~\cite{Mydosh2011, Kung2015, Pourovskiie21,Pourovskii2021,PhysRevLett.121.227003}.   Such hidden subdominant orders or interactions often manifest through the appearance of a so-called missing entropy in thermodynamic measurements~\cite{Ramirez:2003aa,Pomaranski:2013aa,Lau:2006aa,Bareille:2014aa, Tonegawa:2014ab, PhysRevB.75.195113,PhysRevResearch.2.013353, Syzranov:2022aa,Wiebe2007, Santini2009, Mydosh2011, Jeffries2012, Ressouche2012,  Harrison2019, Martiniani2019, Pasztorova2023, chakravarty2001hidden}.   A canonical example includes missing entropy problems arising in frustrated magnetic systems~\cite{Ramirez1994, Ramirez1999, Melko2001, Pomaranski:2013aa, Starykh2015}.

 \begin{figure*}[t]
		 \centerline{\includegraphics[scale=0.90]{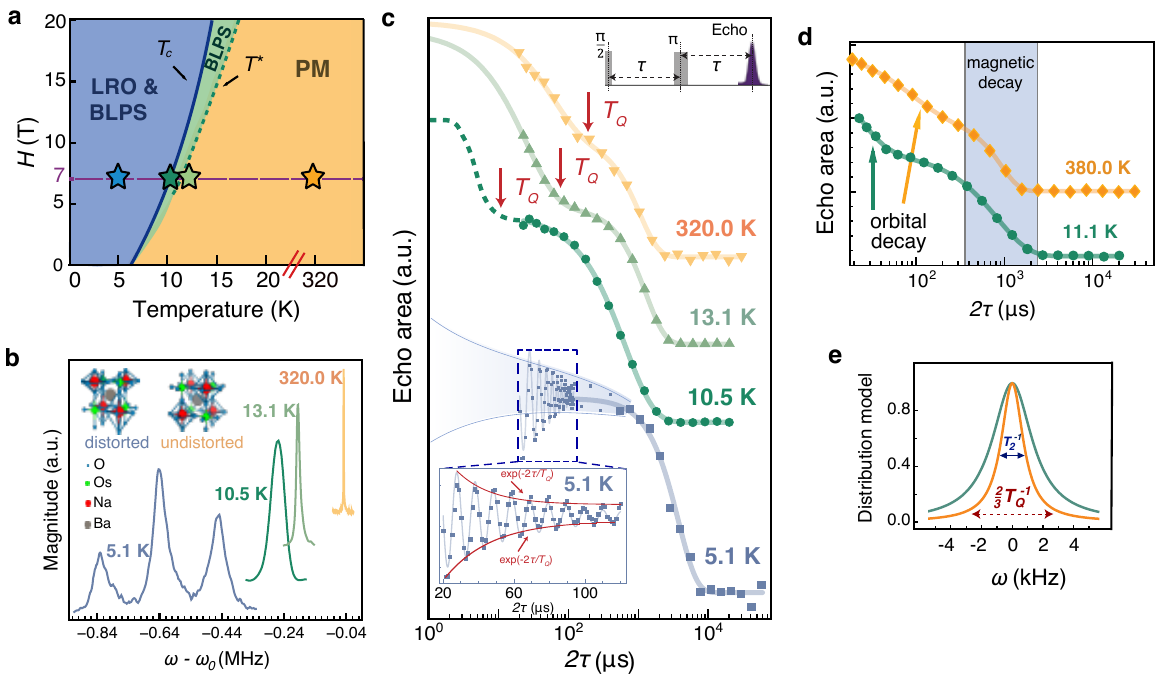}}
			 \vspace*{-0.5cm}
		\caption{\textbf{Orbital order in BNOO inferred from NMR spectra and bi-modal spectroscopy}.
		(\textbf{a}) Illustration of the phase diagram of BNOO. 
        At low temperatures the sample shows both magnetic long-range order (LRO) and broken local point symmetry (BLPS).
        The LRO disappears as the sample undergoes a magnetic phase transition at temperature $T_c$, but the BLPS is still observable in NMR spectra until temperature $T^*$~\cite{Lu2017}.
        Above $T^*$ the system was characterized as a paramagnet (PM) in previous studies~\cite{Lu2017}.
		(\textbf{b}) Representative $^{23}$Na spectra in different phases at the temperatures, denoted by stars in the phase diagram (\textbf{a}),  and a sketch of the local distortions that give rise to a finite electric field gradient (EFG) (upper inset) \cite{Liu2018a,Cong2019}.
        Only one quadrupolar triplet from the two distinct magnetic sites is shown at 5.1~K for clarity.
        The zero  (reference) frequency is defined by $\omega_0 = {}^{23}\gamma H$ of Na nuclei, while the magnetic field is $\vec{H}=7$~T applied parallel to the [001] axis.  
		(\textbf{c}) Spin-echo magnitude as a function of inter-pulse times ($2 \tau$, with echo sequence in upper inset). 
		The 5.1~K decay curve is divided into two sections from two different measurements, an attenuating modulation (lower left inset) and an exponential decay.
		The higher-temperature curves reveal two relaxation processes at different time scales, characterized by a ``plateau'' at time $T_Q$. Error bars are smaller than the symbols used.
		(\textbf{d})
Solid symbols denote  the spin-echo magnitude, measured from the integrated area of the signal in the frequency domain, as a function of the inter-pulse times~$(2\tau)$. Solid lines are fits to the theoretical model~\cite{Carr2022}. The position of the plateau corresponds to the rate~$T_Q^{-1}$. The site frequencies are given by the Lorentzian distribution model shown in~(\textbf{e}).  }  
        \label{fig:Fig1_phaseDiagram}
         \vspace*{-0.5cm}
	\end{figure*}

        Here, we detect an {\em orbital glass} phase that  evolves into an orbital nematic state  on lowering  temperature  
        and  solves the outstanding missing entropy problem  in a Mott insulator with strong SOC, Ba$_2$NaOsO$_6$ (BNOO)~\cite{Erickson2007, Lu2017}.  
       This  nematic order drives unconventional magnetism in this supposedly nonmagnetic compound. 
        The discovery  of the novel orbital phases  was achieved  by directly  identifying 
         the distribution of an orbital order parameter defined by spatial variations of orbital  distortions. 
        The orbital glass ground state is characterized by 
        a spatially distributed  orbital order parameter 
        whose average value is zero. 
         Even though,  this short-range orbital glass ~\cite{Mehran1983,  Tsurkan2005, Fichtl2005,  Mitsumoto2020} persists 
        up to at least 380~K,   its zero mean-value explains why all previous experiments have failed to observe it       
        ~\cite{Erickson2007,  Lu2017,Liu2018, Willa2019,CruzPinhaBarbosa2022}~(Fig.~\ref{fig:Fig1_phaseDiagram}a). 	
      Specifically, we identify  the orbital order, and its associated crystal distortions that generate finite local electrical field gradient (EFG) by way of the quadrupolar interaction between EFG and $^{23}$Na nuclear spin, $I^2_z$   (see Methods). 
       We find that the orbital order parameter has a zero mean in the high temperature $(T)$ disordered/glassy phase while its variance increases upon lowering $T$, reaching the maximum value in the vicinity of the transition temperature ($T_{c}$) into the long-range order (LRO) magnetic state.  
         As the maximum variance    is crossed on further lowering the temperature, $T \lesssim T_{c}$,    the mean of the orbital order parameter acquires a finite value. 
        This is deduced from the observation of a finite EFG that has a global  well-defined principle axis,   
        defining the nematic.   
           Below  $T_{c}$, magnetic LRO develops from this nematic state.  
     This    is a  particular  case  where one of multiply coupled DOFs acquires LRO (magnetic) as a result of the anisotropic spatial variation of another (orbital)~\cite{Mitsumoto2020}.  
 We establish that the remarkable consequence of strong SOC is to give rise to a nematic orbital state that acquires a directional order but preserves global spatial disorder of magnitude and/or anisotropy,  reflected in the variance of the orbital order parameter,   and acts as a  field that induces LRO magnetism~\cite{TopologNem93,Fradkin10, Baek15,PhysRevB.93.125138,Freelon21,Chen:2024aa}.

To detect the  orbital glass and  nematic  states, we implement a novel approach that   concurrently and independently  measures magnetic and orbital order parameters, as well as their distributions, which enables  an unprecedented level of  understanding of ground states of quantum materials with many plausible ordering channels~\cite{Carr2022}, as depicted in \mbox{Figs. \ref{fig:Fig1_phaseDiagram} a \& b}.  
  This bi-modal approach is based on  nuclear magnetic resonance (NMR) spin-echo technique. 
  Here,  for the first time,          we use this method   to measure the  full  spatial distribution of an order parameter without contamination from other coupled DOFs. 
     In essence,   we utilize nuclear spins as local interferometers to perform minimally invasive, phase sensitive measurements of electronic ground states. This technique is in stark contrast to  conventional spectral measurements of matter, which measure the rate of  relaxation 
  caused by quasi-particle excitations.
The local microscopic  Hamiltonian, expressing  weak coupling between nuclear spins $(I)$ and electronic DOFs, is given by
   \begin{equation}
   \label{HamEq}
   \mathcal{H} = \mathcal{H}_Z (I_{z}) +  \mathcal{H}_Q (I_{z}^{2}) + \mathcal{H}_{NH}\, .
    \end{equation}
  The first term (Zeeman) captures magnetic interactions linear in $I_{z}$, the second those that couple to $I^2_z$ via quadrupolar interactions, and  the last denotes all dissipative interactions (see Methods for details). 
  On a fundamental level,  this new method directly probes the  nuclear spin symmetry as a result  of rotating different parts of this local  Hamiltonian. The action of the  nuclear spin  rotation operator is attained by a specific sequence of radio frequency pulses, resonantly and coherently acting on a nuclear spin ensemble.  
More precisely, our pulse sequence is tailored to selectively refocus (tuning constructive/destructive interference conditions) the time evolution of the spins, \ie even and odd spin operators in 
 \mbox{Eq. \ref{HamEq}}, $I_z$ and $I^2_z$, respectively.  Specifically, we refocuses one channel (magnetic) while leaving the other (orbital, which is time-reversal invariant) unfocused, and thus helps  suppress magnetic inhomogeneities   affecting  the spins. The technique can even resolve the distribution of an order parameter that spatially averages to  zero~\cite{Carr2022}. 
    Standard scattering or resonance techniques  are ineffective in resolving zero-centered order parameter distributions, especially in the presence of magnetic inhomogeneities, or with a large  variance  associated with other coupled DOFs. 

We apply this technique to study numerous remaining enigmas in a complex multi-flavor Mott insulator, BNOO ~\cite{Erickson2007,Lu2017,Liu2018a,Cong2019,Willa2019, Chen:2024aa}.  
These  puzzles  clearly indicate the limitations of standard techniques 
  to provide information  on the  nature of the    ground state in   materials with multiple strongly interacting DOFs.   BNOO  consists of  $5d^1$ electrons that have a total angular moment, $J=3/2$, corresponding to four degenerate states. 
    Theory suggests that potential distortions of the crystalline and electronic structure can introduce a multipole term in the  electronic Hamiltonian of the form ${J_z}^n$ ($n$ even), which reduces the number of degenerate electronic spin states from four to two (effective $J=1/2$)~\cite{Goodenough1968, Erickson2007}.
    In this situation, a low-temperature phase transition to magnetic ordering (one ground state, effective $J=0$) usually occurs in one of three distinct ways~\cite{Khomskii2014}:
    a pure magnetic transition with one critical temperature $T_c$;
    the emergence of an octopolar term ($n=4$), where both the orbital and magnetic ordering lift the degeneracy at the same $T_c$;
    an orbital transition caused by a quadrupolar term ($n=2$) followed by magnetic orderings at two distinct temperatures, first in the quadrupolar channel, and then in the magnetic channel~\cite{Khomskii2014}. 
  
\begin{figure*}[t]
	 \centerline{\includegraphics[scale=0.45]{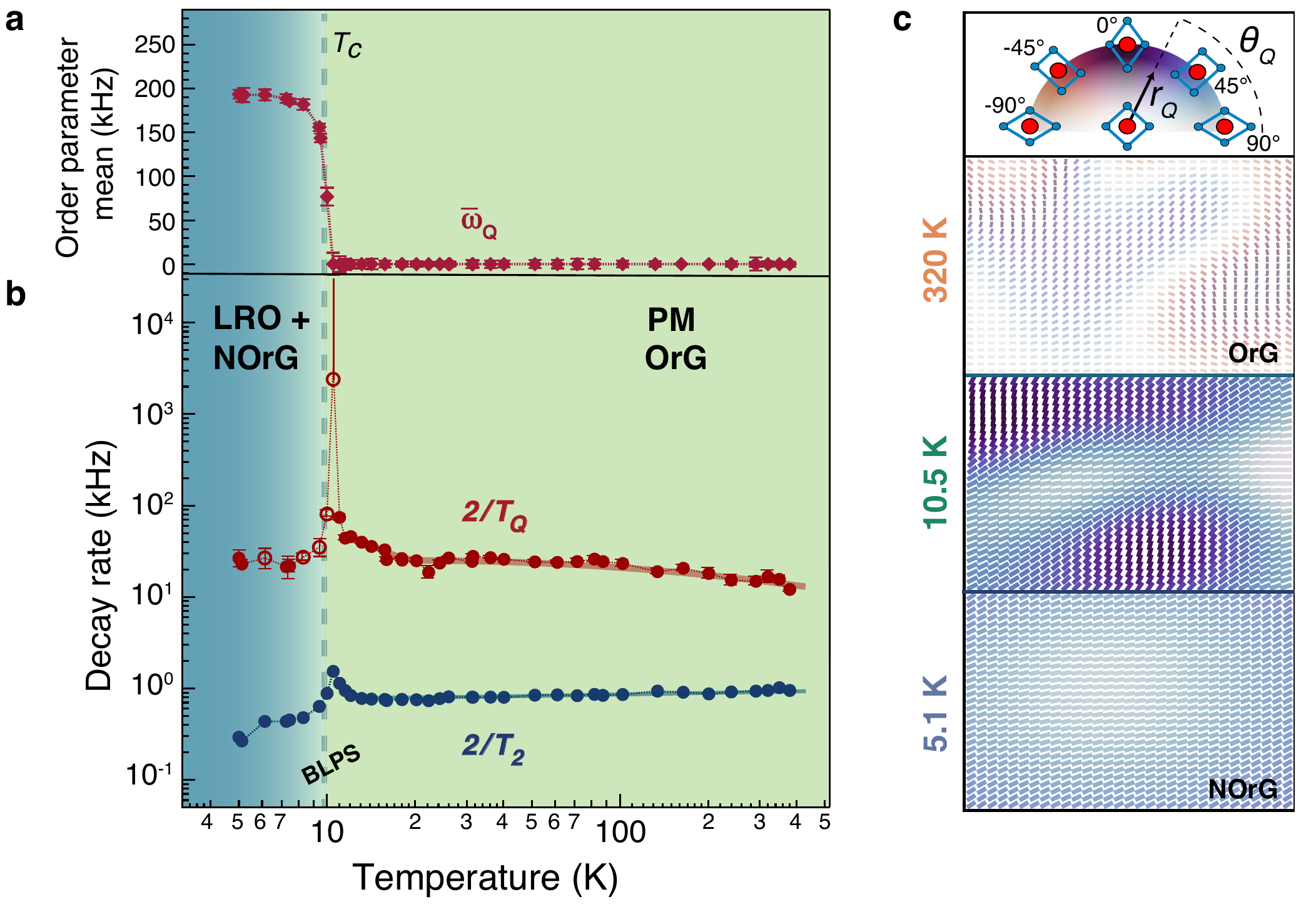}}
	 \vspace*{-0.4cm}
	\caption{\textbf{Order parameters and relaxation processes in BNOO}. (\textbf{a}) 
	$T$ dependence of the orbital order parameter mean,~$\overline{\omega}_{Q}$,  (\textbf{a})  and its distribution  $(2/T_Q)$  (\textbf{b}).     
	 	(\textbf{b})  The magnetic  $(2/T_2)$ and  orbital  $(2/T_Q)$ distributions causing dephasing, as a function of 
 temperature up to 380~K.   The dashed  line denotes $T_c$, the transition temperature below which the LRO magnetic state is formed.  
  The orbital glass state,    denoted as OrG, is characterized by the dominant distribution of the orbital order parameter ~$(2/T_Q)$. A nematic orbital glass (NOrG) is stabilized in the LRO magnetic state. 
 All data were measured in a 7~T~magnetic field  applied parallel to the [001] crystalline axis.  When not discernible, error bars are smaller than the symbols used. The open red points are upper estimates using the spectral linewidth (See Suppl.). Thick shaded red and navy lines serve as a guide to the eye. 
(\textbf{c}) The top panel displays possible in-plane distortions of BNOO's unit cell.
        The angle of the distortion is given by $\theta_Q$ and its magnitude by $r_Q$, yielding an order parameter of the form $\vec{\omega}_Q \equiv r_Q (\sin \theta_Q, |\cos \theta_Q|)$.     The following panels illustrate spatial variations in the BLPS phase above $T_c$ and a global BLPS phase below, as inferred by this work.
        Each unit cell is represented by a colored rod to indicate the local BLPS structure.
        The angle of the rod and its hue indicate the local $\theta_Q$, while the rod's length and its opacity indicate the local $r_Q$. For  visualization purposes, we only show variations in a 2D plane, but in practice the variations should occur in 3D.
    }
    \label{Fig3_Decay}
     \vspace*{-0.5cm}
\end{figure*}

    In BNOO, while a magnetic transition was visible at $T_c$ (Fig.~\ref{fig:Fig1_phaseDiagram}a), no evidence of a highly sought-after, higher-$T$ quadrupolar phase transition has yet been observed~\cite{Erickson2007,  Lu2017, Willa2019, CruzPinhaBarbosa2022}.
	The natural argument is that BNOO must fall into one of the first two transition types, either a non-orbital or simultaneous magnetic and orbital transition.
        Conventional successive symmetry breaking due to orbital order has since been observed in other 5$d^1$ and 5$d^2$ double perovskites~\cite{Takayama2021,Hirai2020,lovesey2021magnetic, tehrani2021untangling,ishikawa2021phase,hirai2019successive}, but its role and nature  in BNOO has remained a mystery.
                Nevertheless, specific-heat measurements~\cite{Erickson2007, Willa2019} found that the entropy loss associated with the magnetic phase transition was consistent with the breaking of a two-fold electronic degeneracy ($J=1/2$) instead of four-fold ($J=3/2$)  -- the missing entropy alluded to above.
		    Our work solves this mystery by revealing that the   entropy is hidden in the configurational entropy of the orbital glass phase, characterized by  a spatially distributed  zero-mean orbital order parameter, as illustrated in \mbox{Fig. \ref{Fig3_Decay}d}. 		     

 {\bf Mysterious nature of high-\textit{T} phase in BNOO.} 
 A paramagnetic (PM) broken local point symmetry (BLPS) phase, characterized by local crystalline distortions in BNOO, as evidenced by conventional   NMR, has only  been observed in the vicinity of the magnetic transition for  $T \gtrsim  T_c$~\cite{Lu2017, Liu2018a}. It was  understood that a multipolar ordering of the electronic structure induces BLPS~\cite{Lu2017,CruzPinhaBarbosa2022}. 
 The sequential onset of BLPS and then the LRO magnetic state   are shown in the 
  field-temperature phase diagram in Fig.~\ref{fig:Fig1_phaseDiagram}a (compiled from references~\cite{Lu2017, Liu2018, Liu2018a, Cong2019}). 
The onset of BLPS was thought to occur in the PM state at a temperature~$T^*$, which does not denote 
a true phase transition~\cite{Lu2017, Liu2018a, Willa2019}.  
 The  NMR spectra of  $^{23}$Na  ($I=3/2$), depicted in~Fig.~\ref{fig:Fig1_phaseDiagram}b,
reveal the distribution of  the hyperfine fields,  induced by local magnetism,     and the electronic charge   distribution (reflecting orbital order and/or lattice symmetry).  A nuclear spin $I =3/2$  interacts via the quadrupolar interaction with a non-zero EFG 
  (see Methods). At sites with cubic point symmetry, the EFG is zero, leading to vanishing quadrupolar interactions. A finite EFG and thus quadrupolar interaction is generated by both electronic and atomic distortions~\cite{Lu2017,Liu2018a,Cong2019}. 
This interaction,  coupling the $I^2_z$ term with the magnitude of the EFG, 
 splits the otherwise single NMR line into $2I$ lines, as illustrated in~Fig.~\ref{fig:Fig1_phaseDiagram}b at  ~$T =  5.1$~K.  However, for small finite values of the EFG, the three peaks are not discernible in conventional NMR spectra and  only significant line broadening is observable (\textit{e.g.}, $T=$~10.5 K). In this case, the  broadening associated with the EFG, \ie charge distribution, cannot be disentangled from that due to the  magnetic DOFs, precluding gaining any further insight into the nature of BLPS above $T_{c}$ by conventional techniques~\cite{Lu2017,CruzPinhaBarbosa2022}. 
 To gain full understanding of  both the missing magnetic entropy at $T = T_c$ and, more generally,  the microscopic relationships between orbital and magnetic orders in BNOO   ~\cite{Chen:2024aa, OrbitPhysTMO,Multipolarhidden19,Voleti2022}, we rely on the capability of the novel 
   bi-modal phase sensitive technique~\cite{Carr2022} to tune two distinct spin evolution times.

{\bf Novel probes of nature of the high-\textit{T} phase.} 
Theoretical treatment of the spin dynamics predicts that spin terms of even power in the nuclear Hamiltonian \mbox{(Eq. \ref{HamEq})} induce a multi-modal relaxation as one varies the time-delay between consecutive pulses~($\tau$)~\cite{Carr2022} (\mbox{Figs. \ref{fig:Fig1_phaseDiagram} c-e}).
Whenever the strength of an even-power term in the Hamiltonian follows a distribution, the spin-spin relaxation will partially attenuate at a rate $T_Q^{-1}$   that is proportional to the linewidth of that distribution.
The critical precession time ($T_Q$) indicates when the quadrupolar-dependent part of the magnetic signal dephases.
This is followed by a second decay in the remaining quadrupolar-independent echo amplitude, caused by conventional magnetic decoherence on the time-scale $T_2$, as depicted in \mbox{Figs. \ref{fig:Fig1_phaseDiagram} c-e} and described by \mbox{Eq. \ref{DecayEq}} in the Methods. Performing a $T_2$-style relaxation measurement at 5.1~K, in the LRO phase (at   $T < T_c$), reveals a sinusoidal modulation in the spin-echo caused by the quadrupolar $(\omega_Q)$ interaction (lower inset of Fig.~\ref{fig:Fig1_phaseDiagram}c), an effect first observed by Abe \textit{et al}.~\cite{Abe1964}, and others~\cite{vachon_133_2006, tokunaga_nmr_2006}.
These oscillations are attenuated (red envelope of Fig.~\ref{fig:Fig1_phaseDiagram}c) on a time scale $T_Q$. The distribution of $\omega_Q$ is centered at a finite frequency \wqo, which  is determined  from the spectral splitting and from the frequency in the sinusoidal oscillation (respectively $190$~kHz and $184$~kHz).
The relationship between the time $T_Q$ and the parameters of the microscopic Hamiltonian is detailed in the Methods.
To sum up, we define the mean of orbital order parameter as   the mean of the distribution of  the  quadrupolar interaction term  $(\overline{\omega}_{Q})$, while the full-width at half-maximum (FWHM) of its Lorentzian distribution is determined by   $2/(3T_Q)$. The
temperature dependence of the mean of the orbital order parameter  $(\overline{\omega}_{Q})$ and the FWHM of its  distribution are plotted in 
 \mbox{Fig. \ref{Fig3_Decay}}.

 A schematic for the formation of  the BLPS sites, \ie  orbital glass,  across the sample is given in Fig.~\ref{Fig3_Decay}c. 
 Even when the magnitude of the crystal distortions are uniform throughout the sample, variations in the angle of the distortion relative to a global $z$-axis set by the experiment can cause variations in the effective scalar interaction strength $\omega_Q$.
    We find that the scalar order parameter has a non-trivial distribution at all measured temperatures, characterized by its  FWHM.
    We directly measure the FWHM from the characteristic decay rate ($T_Q^{-1}$) of the NMR  signal.

 Next, we analyze the temperature dependence of the distribution of the orbital order parameter $\omega_Q$, parameterized by both its mean, \wqo,  and variance, $2T_Q^{-1}$  (see Methods). Results of this analysis are plotted  in~Fig.~\ref{Fig3_Decay}a-b. 
We infer that the mean of the orbital order parameter, \wqo, is zero at all $T > T_c$. On further lowering   the temperature $(T \lesssim T_{c})$, the orbital order parameter acquires a finite mean. We point out   that  \wqo\, characterizes a `hidden'/subdominant order parameter, while thermodynamic measurements are 
dominated by  a response to the magnetic order parameter~\cite{Erickson2007}. 
 Therefore,   the $T$ evolution of   \wqo\, should not be used to  infer  the order of the phase transition and/or compare it to findings from specific heat measurements.  
The rate $T^{-1}_Q$ is largely temperature independent, with only a noticeable growth on approaching $T_c$, \ie when entering the BLPS inter-phase region ($T_c < T < T^*$). Specifically, we  note   that the variance, $2/T_Q$, of  ${\omega}_{Q}$,  becomes larger as the temperature approaches $T_c$ from above, but $2/T_Q$ becomes small again below $T_c$.  Moreover, the magnitude and growth rate of the quadrupolar parameter variations ($T_Q^{-1}$) are much larger than those of the magnetic decoherence processes ($T_2^{-1}$), as shown in \mbox{Fig.~\ref{Fig3_Decay}b}. 
This serves as our key evidence that  the orbital DOF 
 drive magnetic LRO. 
We now proceed to discuss two important implications for the BLPS phase~\cite{Lu2017}: the nature of the short range order above $T_c$ and its role in stabilizing the magnetic LRO at low temperature.

{\bf Orbital glass phase.} 
 The lack of oscillations in the bi-modal spectra bove \mbox{$T_c\sim 10$~K} (Fig.~\ref{fig:Fig1_phaseDiagram}c) implies that crystalline distortions exist, but no specific direction is preferred, leading to $\overline{\omega}_{Q} = 0$ as shown in \mbox{Fig. \ref{Fig3_Decay}a}. 
If the distortions \textit{only} occurred along specific facets of the crystal, with sharp transitions between each orientation domain, a handful of unique values of $\omega_Q$ would be present at a given $T$.
In this case, a finite number of frequencies would be visible in the spectral transform of the $\tau$-decay curves, but we observe a decay which is well described by a Lorentzian distribution of a width proportional to~$T_Q^{-1}$.
We hypothesize that the system is composed of domains which may prefer specific facets, but that the regions between them are smooth and have slowly varying transitions between distortion angles. While the nature of the spatial variations is predominantly angular, it is possible that there are some weak variations in the magnitude of the crystalline distortions as well. 
Strong variations in  the magnitude would couple to  the accessible energy of the phonons, which can lead to a strong temperature-dependence of $T_Q^{-1}$ at higher temperatures.  Nevertheless, modulations in the magnitude can combine with the glassy variation in the distortion orientation to create a nearly perfect Lorentzian distribution of local $\omega_Q$ values across the entire sample.
This distribution of the orbital ordering is  described as an \textit{orbital glass}. 

By probing the local spin precession, our bi-modal technique is sensitive to fluctuations from $\mu$s to ms timescales, allowing us to rule out dynamical fluctuations across these scales. 
Thus, our measurements establish that the quadrupolar distribution is caused by static variations rather than dynamic fluctuations. 
This is inferred from our ability to probe dynamical effects on multiple timescales ranging from slow ($>1$~ms) to fast ($1~\mu$s) and the lack of any significant  $T$ 
 dependence of quadrupolar distribution $(T_Q^{-1})$, away from $T_{c}$ (see the Methods for more details). 

Insight into the orbital glass is gained by noting three key properties of both the orbital~($T^{-1}_Q$) and magnetic~($T^{-1}_2$) distribution variances (noise) near the phase transition (Fig.~\ref{Fig3_Decay}):
(i) while the magnetic noise reduces by nearly an order of magnitude below the transition into LRO, the value of the rate $T^{-1}_Q$ at $5.1$~K is identical to that at $50$~K;
(ii) in the BLPS inter-phase region, the relative growth of the distribution of orbital order parameter, \ie distribution of  distortions, is much larger than that of the magnetic variations, $\propto T^{-1}_2$;
(iii) near the phase transition $T \sim T_c$, the width of the orbital glass distribution, FWHM $\propto T^{-1}_Q$,  
 diverges significantly faster than  $T^{-1}_2$ and well beyond its value on either side of the transition. 
These three observations, taken together imply  the BLPS distortions  stabilize the magnetic fluctuations, ultimately driving the phase transition to a magnetic LRO.
 As the temperature is lowered to $T_c$, the variance of the orbital distortions  grows  over an order of magnitude,    while those in  the magnetic channel increase by less than a factor of two (\mbox{Fig. \ref{Fig3_Decay}b}).
                         Then, just after reaching the critical temperature $T_c$ $(T<T_{c})$, the magnetic variations decrease to well below their high temperature value, while the orbital ones  
 remain similar to their high temperature values.

 {\bf Low-$T$ orbital nematic state.}   
The observation of comparable  orbital variations below and above $T_c$ demonstrate that orbital glass persists in the LRO magnetic state.  The detection of a finite EFG, evident by the presence of  oscillations
in the bi-modal spectra,  below  $T_c$  reveals that the sole effect of the spin ordering on the orbital DOF is to impose  preferred direction of distortions, which induced a finite \wqo, as depicted  in \mbox{Fig.~\ref{Fig3_Decay}a}.  These conclusions provide very stringent symmetry-based constraints on possible microscopic models for systems with complex  multipolar interactions. 
More importantly,  the emergence of a  preferred direction of the EFG below  $T_c$ reveals that  the low-$T$ magnetically ordered state  coexists with an 
{\it orbital nematic phase} (the nematic-like liquid-crystal state of the orbitals)   characterized by directional order, which preserves global  spatial disorder, \ie glassiness. 
Specifically, while the direction of the nematic order is well defined by the principle axis of the EFG,  the magnitude and/or anisotropy  of its order parameter remains  spatially disordered as in a glass, as illustrated in \mbox{Fig. \ref{Fig3_Decay}b \& c}.  %
This is     a peculiar, purely  orbital nematic state without accompanying  
spin fluctuations~\cite{Fradkin10, Freelon21}. 
By nature of our measurement, we  directly observe this nematic orbital state   without any variation of an external field (\textit{e.g.}, light, strain)~\cite{Shimojima:2019aa,PhysRevB.99.180102}. 

{\bf Relationship between orbital nematic  and LRO magnetism.}   
 To gain further insight into the nature of the orbital nematic state and its relationship to  magnetism, we perform a careful examination of the temperature evolution of the NMR spectra~\cite{Lu2017, Liu2018}  and \wqo\, 
(\mbox{Figs. \ref{Fig3_Decay}a}).  
At high temperatures, a single peak is seen in the spectrum. At a lower temperatures below 10 K, the peak  splits into three. This can be understood as quadrupolar splitting due to finite  EFG and suggests the formation of nematic order. A striking feature of the data is the emergence of another triplet at still lower temperatures. We ascribe this feature to magnetic ordering. Indeed, all crystallographically equivalent sites are equivalent magnetically in the absence of magnetic order, but two types of inequivalent sites are present in a magnetically ordered system. It is thus natural to conclude that nematic order drives a transition to a magnetically ordered state.
We deduce   that  the onset of the orbital nematic state precedes that into the LRO magnetism, because spectral signature of finite EFG is observed at $T >T_{c}$ ~\cite{Lu2017, Liu2018}. 
 Therefore, it is likely that two consecutive   phase transitions occur in BNOO on  decreasing  $T$.  
  First, the static disorder in the orbital glass phase induces an orbital nematic state. The nematic, characterized by a finite \wqo,  acts as a  uniaxial random field that drives the ferromagnetic ordering in the plane transverse to the nematic axis at lower $T$ ~\cite{Lu2017, Liu2018}. In BNOO,   the orbital nematic    induced    anisotropy, along the [001] crystalline axis ~\cite{Lu2017, Liu2018}, reduces magnetic fluctuations   and thus promotes magnetic ordering. 
    This resulting LRO magnetic state is  a peculiar two sub-lattice ferromagnet whose spins are aligned in the [110] plane~\cite{Lu2017}.  
 Evidently, the significant SOC in BNOO induces a more complex transverse magnetic ordering compared to that in pure magnetic systems ~\cite{Chudnovsky86,Feldman98}.

\begin{figure}[t]
 \centerline{\includegraphics[scale=0.45]{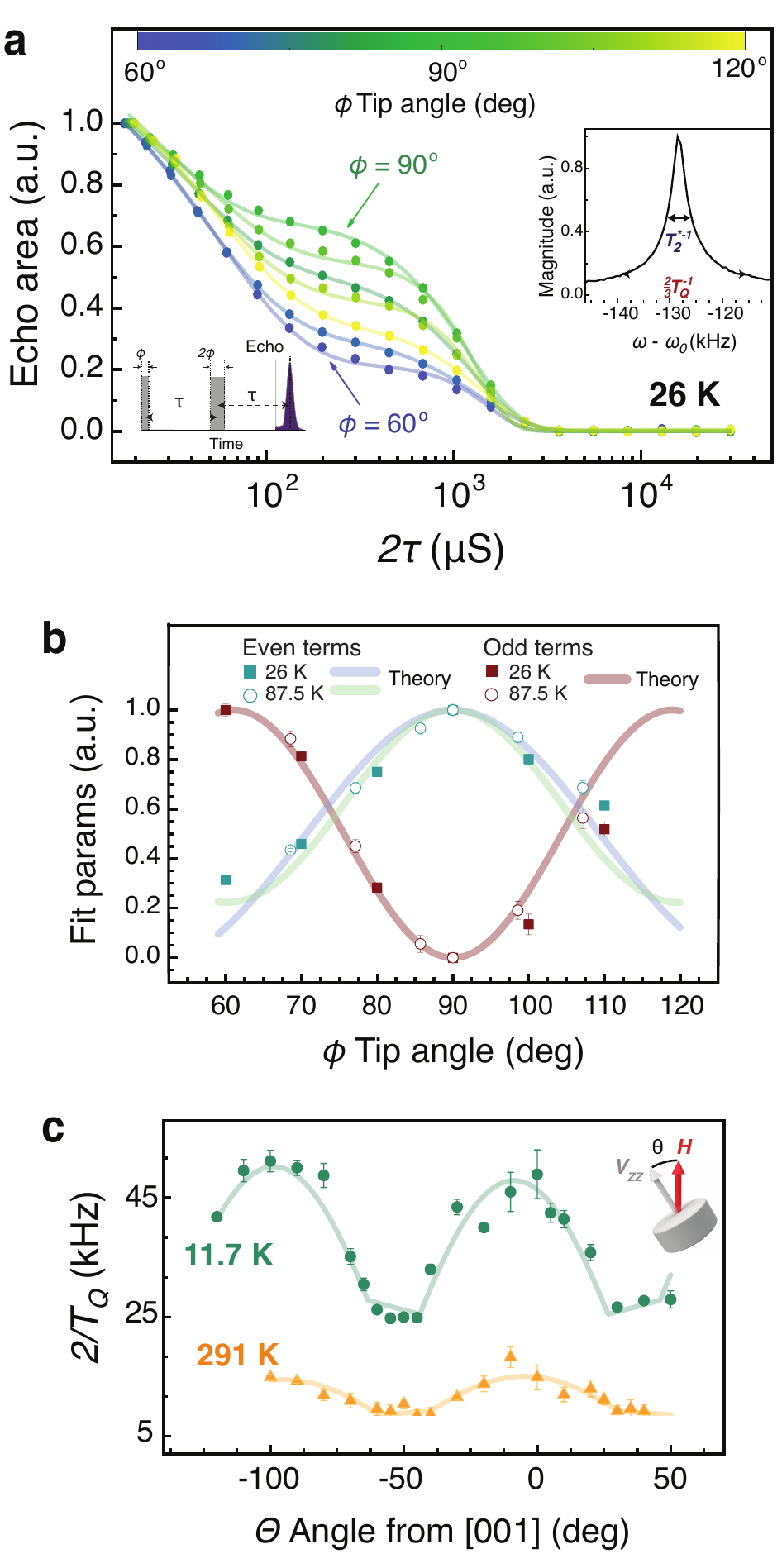}}
	 \vspace*{-0.4cm}
	\caption{\textbf{Angle and pulse dependence measurements}. 
    (\textbf{a}) 
    Echo area as a function of the delay time $2\tau$ at different tip angles $\phi$, at 26~K, fit with Eq.~\ref{eq:I_ABC}. Here, we take $A=C$ as they are close in value around the perfect $\pi$/2 pulse~\cite{Carr2022}. The average value \wqo\, is zero. 
     (\textbf{b})  Tip-angle ($\phi$) dependence of the prefactors in Eq.~\ref{eq:I_ABC}: $A, C$ (associated with exponents of $2\tau$ and $0$, labeled even) and $B$ (associated with $\tau$, labeled odd). Data is extracted from the fits given in (\textbf{a}), with one standard deviation given by the error bars. Solid lines are theoretical predictions~\cite{Carr2022}.
    (\textbf{C}) Dependence of the rate $2/T_Q$ on the orientation of the sample~($\theta$) between its principal axis $V_{ZZ}$ (local EFG) and the applied field \textbf{\textit{$H$}} = 7~T, at two temperatures. Solid lines are a fit to  Eq.~\ref{eq:quadrAngle}, implying a distribution of quadrupole moments. 
	 When not discernible, error bars are smaller than the symbols used. }
	\label{Fig2_knobs}
	 \vspace*{-0.7cm}
\end{figure}

{\bf Missing entropy elucidated.} We have provided a clear answer to why BNOO has an effective \mbox{$J=1/2$} entropy at the critical temperature $T_c$.
Our measurements lead to the conclusion that the missing entropy is hidden in the configurational entropy of an orbital glass phase and not in dynamical fluctuations of the magnitude or directions of the crystal's distortion. 
The magnetic dephasing $(T_2)^{-1}$ 
are comparable in magnitude to $T_Q^{-1}$ but cannot contribute to the entropy change because of their gradual   $T$-dependence (see Supplementary Materials). 
The missing entropy is therefore distributed over a wide temperature range in the electronic orbital DOF, \ie orbital glassiness. 
 As  missing entropy puzzles  are often associated with hidden-order phases, this technique should be transferable to the study of other challenging magnetic systems, such as those with frustrated geometry~\cite{Wiebe2007, Erickson2007,Tripathi2007, HidOrderFisher15, Starykh2015,Mitsumoto2020,Pourovskiie21,  Ramirez:2003aa,Pomaranski:2013aa,   Melko2001}. 
Frustration suppresses ordinary long-range order and in some cases stabilizes PM states down to ultra-low temperatures. In the presence of orbital degeneracy, a spin-orbital liquid and/or glass states, such as ones that we directly identified in this work, are plausible~\cite{Mitsumoto2020,Samarakoon:2017aa, OrbitPhysTMO}.
Identifying the ground state in frustrated spin systems via measurements of excitations is difficult as there often exists a highly degenerate manifold of possible ground states with similar spectra~\cite{Gingras01, Lee:2002aa, OrbitPhysTMO,Chen:2024aa}.

{\bf Orbital origin of disorder.} 
Our proposed physical picture relies on the supposition that $T_Q^{-1}$ is a measure of the scalar quadrupolar interaction's distribution, which is generated by a distortion-induced EFG at the ${^{23}}$Na sites~\cite{Lu2017, Liu2018a, Cong2019}. To further test this interpretation, and obtain two independent confirmations of the orbital order, we expand our analysis of the bi-modal response to include tip and tilt angles. 
The tip angle is the quantity of net nuclear spin rotation around the applied field given by the strength and duration of the first radio frequency pulse in the echo sequence. Therefore, the tip angle is a knob that controls the efficiency of refocusing of the orbital terms, $I_z^n$ ($n$ even).  The tilt angle is the angle between the principal axis of the EFG and the applied magnetic field, and is achieved in the experiment by physically rotating the sample.

Although the $T_Q$ time scale encodes the most important information about the distribution of $\omega_{Q}$, the tip-angle dependence of the ``plateau'' height allows us to confirm the quadrupolar/orbital nature of the $T_Q$ decay. The relative plateau height as a function of $\phi$   is  extracted in terms of the even and odd terms of the theoretical model (see the tip-angle dependence section in Methods). The tip-angle dependent plateau height shows excellent agreement between theory \cite{Carr2022} and experiment (Fig. \ref{Fig2_knobs}a) confirming that the $T_Q$ is a true measure of the orbital disorder. 

To independently confirm that the nature of the spatial variations is predominantly angular above $T_{c}$, we plot the $T_Q^{-1}$ as a function of the tilt-angle $\theta$ in Fig.~\ref{Fig2_knobs}c.  We note that a pure magnitude variation  
  would be inconsistent with the large offset in $T_Q^{-1}$. 
That is, in addition to a uniaxial magnitude variations, there must be some angular variations as well to account for  the  offset in $T_Q^{-1}$ versus  the til-angle in Fig.~\ref{Fig2_knobs}c. 
The rate $T_Q^{-1}$ follows the angle dependence resulting from the crystalline quadrupolar interaction with two key additional features.
First, $T_Q^{-1}$ close to the magic angle ($\approx 54.7^{\circ}$) (see Methods) is almost constant, which is well captured assuming at least two BLPS domains with differing principal axes (Eq.~\ref{eq:quadrAngle}). In contrast, a uniform orientation of the material's principal axis would result in a sharp minimum of $T_Q^{-1}$ at the magic angle~\cite{Liu2018a}. 
Second, the $\theta$-independent constant offset in $T_Q^{-1}$  rules out dephasing caused by a dipolar interaction because such dephasing has a perfect averaged refocusing at the magic angle (discussed further in the Supplement).

{\bf Discussions.} 
Our demonstration of selective refocusing of local spin terms, based on the symmetry of the spin interaction, can be applied to single spin systems with a high spin number, such as the molecular magnets embedded in solid matrices~\cite{Chiesa:2020aa,Yu:2021aa},  to achieve high fidelity of spin control. 

For strongly correlated spin-orbit materials, our symmetry-selective refocusing allows one to disentangle the role of orbital and magnetic interactions in poorly understood low-temperature phases~\cite{Santini2009, WitczakKrempa2014, Fu2015, Patri2019, Yin:2022aa}. 
 Our observation of orbital nematic-induced magnetic order provides the first material platform to test theoretical models that aim to identify   microscopic mechanisms that drive the emergence of complex phases  in systems with multiple interacting DOF~\cite{Chudnovsky86,PhysRevLett.98.156801}. The main objective of these theories is to establish  a road map for  tuning emergent/hidden  orders by randomness engineering.   
 
\section{Methods}
         \textbf{NMR data acquisition.} A state-of-the-art, custom-built NMR spectrometer with advanced quadrature detection capabilities in combination with a high-homogeneity superconducting 7~T magnet at Brown University is used. The temperature control above room temperature (roughly 294 K) is achieved using a heater around the sample chamber, while cooling below room temperature is performed using a helium variable temperature inset.
	
 \bigskip
 
	   \textbf{Sample.} We use the same high-quality single BNOO crystal, grown from a  molten hydroxide flux, described and studied in previous work published in Ref.~\onlinecite{Lu2017}. 
	   Based on clear, sharp peaks in the NMR spectra, we are confident about the sample quality. 
	   The sample is aligned such that its [001] crystalline axis is parallel to the magnetic field, which in this sample coincides with the principal axis of the electric field gradient (EFG)~\cite{Lu2017}.
	
 \bigskip
 
	  \textbf{Competing fluctuations.} The Hamiltonian of the ${}^{23}$Na nuclei in BNOO in the principal axis system (PAS) of the EFG is given by 
	\begin{equation}
    	\begin{split}
    		\label{eq:H}
    		& \mathcal{H} = H_Z + H_Q + H_{NH}, \\ 
    		& H_Z = \omega_0 I_z, \\
    		& H_Q^{PAS}  = \frac{\omega_Q}{2} \big[(3 I_z^{'2} - I^2) + \eta (I_x^{'2} - I_y^{'2}) \big],
    	\end{split}
	\end{equation}
	where $\omega_0 = {}^{23}\gamma H$ is the Larmor precession frequency of ${}^{23}$Na, whose variations (magnetic noise) are characterized by a broadening $\Gamma_Z$, and  $\omega_Q$ is the magnitude of the nuclear quadrupolar interaction caused by a locally non-zero EFG and whose variations are characterized by the width of the distribution of the order parameter (DOP), $\Gamma_Q \equiv \frac{2}{3} T_Q^{-1}$, primed spin operators are defined in the PAS frame, and $\eta$ is the anisotropy. The distribution of quadrupolar interaction strengths, $\omega_Q$, is defined by its full-width at half maximum (FWHM), $\Gamma_Q$, while its center is denoted by \wqo. The different spin operators between the Zeeman and quadrupolar terms of the Hamiltonian indicate that the two corresponding coordinates systems are not \textit{a priori} aligned. The last term, $H_{NH}$, captures  interactions with all other DOFs responsible for dissipation. 
    Because we have restricted our Hamiltonian to a single nuclear site, $H_{NH}$ encompasses all other interactions with the environment, including weak interactions with  nearby nuclei (spin-flips, $I_+, I_-$). 
We are comparing the measured magnetic, spin-spin decoherence $T_2$ and $2/T_2^{*} \equiv \Gamma_Z$  with the quadrupolar dephasing in terms of a characteristic dephasing time, given by $T_Q$. 
 	  
	  The bi-modal  method employed to measure $T_{Q}$ and $T_2$  is related to the spin-spin  
 relaxation time measurement $(T_2)$ and, unlike  the NMR relaxation rate  $(T_1T)^{-1}$, it does not involve energy transfer between nuclear spins and an electronic/phononic bath.

    \bigskip
   \textbf{Spin-echo sequence.}
   Raw data was acquired by signal-averaging  a standard Hahn-echo sequence $\pi/2-\tau-\pi$.     The repetition time between consecutive echos 
  (ranging from around 1.25~s to 8~s) is 5 times larger then $T_1$ at corresponding temperatures to ensure that all nuclear spins have relaxed back to their equilibrium by the time the next single spin-echo sequence  starts. The delay time $\tau$ is modified on the exponential scale and is varied based on temperature from a few $\mu$s to 10's of ms. The $\pi/2$ pulse duration is also temperature dependent, ranging from 1~$\mu$s to 4~$\mu$s. Four-step phase cycling is used to eliminate any magnetic inhomogeneity contribution, $\Gamma_Z$, and enhance the signal-to-noise ratio.
   
    \bigskip
  \textbf{Spin-echo modulation.}
  The low temperature ($\sim$5~K) decay curves of the echo amplitude intensity as a function of $2  \tau$ are divided into two regions. In the small-$\tau$ region, inset of Fig. \ref{fig:Fig1_phaseDiagram}c, there is echo modulation due to the non-zero angular averaged quadrupole coupling constant \wqo $\not = 0$   and the decay is dominated by quadrupole relaxation, $\Gamma_Q$. We fit echo intensity, $I(2\tau)$, to the following function~\cite{Carr2022}:
    \begin{equation}
    \label{DecayEq}
    	I(2\tau) \propto \left( A \cos\left(3 \overline{\omega}_Q 2 \tau\right) e^{-2 \tau/T_Q} + C \right) e^{-(2 \tau/T_2)^{\alpha}}.
    \end{equation}
    For the $T_2$ decay, we use a stretching exponential decay fit. Typically, the stretching exponent $\alpha$ ranges from 1 (exponential decay) to 2 (Gaussian decay).  
    \bigskip
    
    \textbf{Bi-modal relaxation.}
    Above the BLPS temperature ($\sim$10 K), the average quadrupole coupling constant is zero, $\overline{\omega}_{Q} = 0$. The model for the $\tau$-dependent amplitude of the spin-echo then is:
    \begin{equation}
    	\label{eq:I_AC}
    	I(2\tau) \propto \left( A e^{-2 \tau/T_Q} + C \right) e^{-(2 \tau/T_2)^{\alpha}}.
    \end{equation}
 
   \bigskip
   
\textbf{Tip-angle dependence.} The efficiency of the time-reversal pulsing is maximized at a spin tip angle of $\phi = 90^\circ$, which corresponds to a minimization of the BLPS (quadrupolar)-induced dephasing, \mbox{Fig. \ref{Fig2_knobs}c}. We have shown in Ref.~\onlinecite{Carr2022} that whenever the tip angle departs from a perfect $\pi/2$ pulse, the spin-echo decay curves in a sample with a distribution of quadrupolar ordering, assuming \wqo $= 0$, have a form given by :
\begin{equation}
	\label{eq:I_ABC}
	\begin{split}
		I(2\tau,\phi) \propto & \left( A e^{-2 \tau/T_Q} + B e^{-\tau/T_Q} + C \right) e^{-(2\tau/T_2)^{\alpha}}.
	\end{split}
\end{equation}
Here, an extra term $B$ with half the characteristic decay time-scale is needed to capture the tip-angle dependence. In the theoretical model that derives Eq. \ref{eq:I_ABC}, the variables $A$, $B$, and $C$ are all dependent on the magnetization tip-angle $\phi$ and, to good approximation, we can take $A \approx C$ for $|\phi - 90^\circ| < 30^\circ$~\cite{Carr2022}.  Experimentally, we have varied the tip angle from $\phi \approx 60^\circ$ up to \mbox{$\phi \approx 110^\circ$} and fit the curves with Eq.\ \ref{eq:I_ABC} above, \mbox{Fig. \ref{Fig2_knobs}c.}

\bigskip  

\label{angle}
\textbf{Angle dependence.}
    In the lab frame, the quadrupolar Hamiltonian up to first order is given by
 \begin{align}
     H_Q^{LAB}  &= \frac{\omega'_Q (\theta)}{4} (3 I_z^2 - I^2), \\ 
     \omega'_Q(\theta) &= \omega_Q (3\cos^2 \theta - 1 - \eta\cos2\varphi\sin^2\theta)
 \end{align}
 where $\theta/ \varphi$ is the polar/azimuthal angle between the applied field $H$ and the PAS, defined by the principal axes of the EFG. The nuclear magnetization dependence on the angle $\theta$ is thus fit to a form which assumes at least two domains of perpendicular PAS:
\begin{equation}
\begin{split}
\label{eq:quadrAngle}
    I(\theta)&=\left |\frac{I_1}{2}\big[3\cos^2(\theta+\delta)-1\big] \right| \\
   & + \left|\frac{I_2}{2}\big[3\cos^2(\theta+\frac{\pi}{2}+\delta)-1\big] \right|+I_0.
\end{split}
\end{equation}
We take the principal axis of the quadrupole tensor $V_{zz}$ to be parallel to the applied field at $\theta = 0^\circ$, and assume the anisotropy is $\eta = 0$, reflecting a BLPS phase with tetragonal symmetry. For a single phase, there is a ``magic angle'' of $\theta = 54.7^\circ$ at which the signal would go to zero or a constant offset, $I_0$ \mbox{ (Fig.  \ref{Fig2_knobs}c)}. The angle $\delta$ is a fitting parameter to account for slight misalignment in the orientation of the sample relative to the applied field. The sample is rotated on the $ac$ plane from [001] axis to [$\pm$100] axes. \\
 
 \bigskip
 \textbf{Dynamical effects} Our conclusion that    the observed quadrupolar distribution is caused by static variations rather than dynamic fluctuations is based on the following facts.
First, as no evidence of structural distortion has been observed by scattering experiments above $T^*$, we can rule out uniform but slowly varying ($>$1 ms) changes as the origin of the distribution.
We can also rule out dynamical fluctuations on a timescale similar to the typical experiment time ($100~\mu$s), because the plateau appearing in the multi-modal measurement relies on refocusing of time-independent local Hamiltonians.
The dynamical fluctuations cannot be much faster than the typical experiment time, as then the plateau would have occurred before the probed $\tau$ values.
In general, a dynamic origin of fluctuations is unlikely given that $T^{-1}_Q$ is only weakly temperature dependent above $T_c$.
Our measurements do not rule out dynamic \textit{and} spatially inhomogeneous fluctuations which occur on a time scale longer than $T_2$, but those would not contribute to our measurement of $T_Q^{-1}$. \\

\noindent {\bf Acknowledgements.}
We thank   Z. Wang, and K.\ Plumb for helpful discussions.  
 This work was supported by the National Science Foundation under grant No.~OIA-1921199  and grant No.~DMR-1905532 (VFM). A.D.
was partially supported by the U.S. Department of Energy, Office of Science, Office of Basic Energy Sciences, under Award Number DE-SC0022311. \\

\noindent {\bf Author Contributions.}
I.R.F. prepared the samples. R.C.  and  I.K.N.   performed the experiments and basic NMR data analysis. 
I.K.N. and S.C. developed the MMS method and data analysis protocols. 
 I.K.N.,  S.C., A.D., C.R., and V.F.M. analyzed the data and developed data interpretation.  D.E.F. assisted   with data interpretation.  V.F.M. provided conceptual advice  and supervised the project.   I.K.N., A.R., S.C., A.D., C.R., and V.F.M.  wrote the paper. All authors discussed the results, and commented and edited the manuscript. \\


\bibliography{refsOO.bib}


{\end{document}